\documentclass[english]{revtex4}
\usepackage[T1]{fontenc}
\usepackage[latin1]{inputenc}
\usepackage{amsmath}
\usepackage{amssymb}

\makeatletter

\providecommand{\LyX}{L\kern-.1667em\lower.25em\hbox{Y}\kern-.125emX\@}

\usepackage{babel}
\makeatother
\begin{document}

\title{Ballistic dynamics of a convex smooth-wall billiard with finite escape
rate along the boundary }

\author{Igor Rozhkov, Ganpathy Murthy}

\address{Department of Physics and Astronomy, University of Kentucky, Lexington,
Kentucky 40506}

\begin{abstract}
We focus on the problem of an impurity-free billiard with a random
position-dependent boundary coupling to the environment. The response
functions of such an open system can be obtained non-perturbatively
from a supersymmetric generating functional. The derivation of this
functional is based on averaging over the escape rates and results
in a non-linear ballistic $\sigma $-model, characterized by system-specific
parameters. Particular emphasis is placed on the {}``whispering gallery
modes'' as the origin of surface diffusion modes in the limit of
large dimensionless conductance. 
\end{abstract}
\maketitle

\section{Introduction}

Wave billiards with smooth boundary walls find a wide variety of applications
in condensed matter physics \cite{PhysNano}, microwave and acoustic
chaology \cite{Stockmannbook}. For chaotic transport in nano-size
two-dimensional electronic systems such as quantum dots, wires, junctions,
and corrals, billiards serve as convenient theoretical models for
a confined electronic gas. Chaos in a cavity might be attributed to
the presence of impurities or to the properties of the boundary itself.
A review of closed billiard dynamics can be found in \cite{Berry1}.
Convex hard-wall cavities open to the environment are also well studied
within the semiclassical approximation \cite{SCA}.

An intermediate situation arises when one allows for a small but finite
coupling to the outside world all around the boundary. Furthermore,
if this coupling is described by a random set of coupling coefficients
to a large number of ideal leads, the resulting system acquires a
natural statistical description \cite{arxivePaper}. Indeed, in studies
of nanostructures one typically aims at the statistical properties
of various response functions, computed through the cavity Green's
function. In experiments the statistics are obtained via spectral
averaging or from an ensemble of different boundary configurations,
corresponding to the same spectral density. In both regular and chaotic
cavities, the most well studied statistical ensemble is the one defined
over the different configurations of the internal impurities \cite{WhiteBook,Stone1,Disordred,Efetovsbook}.

The notion of a statistical ensemble does not come naturally for the
analytical study of a billiard with no internal scattering potential,
no magnetic impurities or random magnetic field. The scattering, which
in this case takes place exclusively at the boundary, is the source
of stochasticity. In the simplest case the scattering is specular,
but it can also be diffusive, as in \cite{Mirlin1}. Furthermore,
the scattering may be accompanied by the escape of scattered waves
into the exterior \cite{arxivePaper}, as, for example, in quantum
dots \cite{Marcus} or quantum corrals \cite{QuantumCorrals}. In
this paper we present a more detailed description of the method proposed
in \cite{arxivePaper}. The model system we study is a billiard weakly
attached to the large number of ideal infinite leads \cite{arxivePaper}.
For simplicity and clarity of presentation we consider one open channel
in each waveguide, with a coupling coefficient which is Gaussian distributed
around a small, but non-zero mean. The mean and the width of the coupling
coefficient distribution are assumed to vary smoothly with the position
of the lead on the perimeter.

For this dot, we construct the supersymmetric generating functional
\cite{WhiteBook,Efetovsbook} and performing the average over the
ensemble of realizations of coupling coefficients, we obtain a {}``surface''
non-linear $\sigma $-model (NL$\sigma $M). Its {}``diffusion''
modes are confined to the boundary of the dot. In convex nearly closed
billiards these correspond to the so called whispering gallery modes
(WGM), which are high angular momentum modes corresponding to the
classical trajectories running alongside the billiard walls. The WGM
are exponentially less likely to escape compared to the modes with
incidence directions close to the lead normals, as can be inferred,
for example, from \cite{ScweitersAlfordDelos}. Thus, the response
functions at very large times are expected to be dominated by these
modes, and hence can be non-perturbatively calculated from our supersymmetric
functional.

Some of the technical tools we employ, such as the non-Hermitian effective
Hamiltonian have been frequently used along with random matrix theory
(RMT) \cite{Scattering,VWZ,NonHermitianRM}. The use of RMT implies
a restriction to the universal regime. Here we pursue a description
of the nonuniversal regime of the billiard dynamics, thus covering
a much broader energy range than is possible with RMT. We are also
interested in spatially dependent characteristics of the system, inaccessible
via RMT. 

The assumption of only one open channel is not essential. The generalization
of our approach to an arbitrary number of open channels is straightforward.
The choice of random Gaussian distribution of the channel strengths
is in fact quite realistic; the actual values of couplings (and the
number of open channels) are not known in general \cite{Lobkis}.
It is usually argued, at least in situations where the RMT is applicable,
that quantitative results do not depend on the coupling being random
or constant \cite{Scattering}. For our development, however, the
randomness is essential as we briefly explained above.

We note that our model applies, \emph{mutatis mutandis}, to a system
with local or extended sources of damping \cite{Lobkis,LossySystems}.
Besides quantum dots \cite{Marcus}, there are other mesoscopic billiards
which fall into this category, such as quantum and optical corrals
\cite{QuantumCorrals,OpticalCorrals}, optical resonant cavities \cite{NockelStone},
and the artificial atoms proposed in \cite{ResonantMagneticVortices}.
Here also we expect the WGM to play an important role in the long-time
Green's functions. The dependence of escape rate on the angular momentum,
important in both optical and electronic systems, is also incorporated
in our model.

Finally, weak interactions in nanostructures with large dimensionless
conductance creates an additional motivation for our study. For a
ballistic cavity, in which electronic interactions are of interest,
while the shape is not, the convex smooth-wall billiard represents
a sufficient starting point. It might be possible to use the results
of our present analysis in the large-$N$ approach of \cite{longpaper}
to the interacting ballistic case. We also note in passing that the
fundamental problem of constructing a $\sigma $ model for a closed,
impurity free ballistic billiard has proven to be technically challenging
\cite{Mirlin1,MKhm,AleinerLarkin,EfetovBallistic}. Our work shows
that a random coupling to the environment acts as a natural regularizer
and allows us to circumvent many of the technical difficulties.

The plan of this paper is as follows. Section II reviews the procedure
for {}``integrating out'' the leads to arrive at a non-Hermitian
Hamiltonian which describes electronic scattering in a quantum dot.
In Section III, we construct the supersymmetric generating functional
for the correlation functions and carry out the ensemble average,
introducing a supermatrix field for each lead to decouple supervector
fields. The continuous model resulting in the limit of large leads
number is analyzed in Section IV. We demonstrate how a Born-Oppenheimer-like
approximation \cite{Zaitsev1} can be employed in the solution of
the effective problem, which results from the saddle point evaluation
of superintegral. This is done in Section V. We complete the derivation
of the surface diffusion NL$\sigma $M in Section VI and present our
summary in Section VII.

\section{Effective non-Hermitian Hamiltonian}

It is common in mesoscopic physics to study open systems using their
closed counterparts and vice versa. The eigenfunction structure of
closed dots affects the transport properties of the open dots, while
the leads attached to the dot can change the nature of the dynamics
from regular to chaotic \cite{OpenSystems}. In the RMT context, the
scattering approach \cite{Scattering} to mesoscopic billiards relies
on the decomposition of the system into internal and external subparts.
The internal is described by $N$ bound states, while the external,
by $M$ channel states propagating along the hard wall leads to infinity.
To set up the analysis one introduces (i) the $N\times N$ internal
Hamiltonian $H_{in}$, and (ii) $M\times M$ $S$-matrix, which relates
incoming and outgoing wave amplitudes in the asymptotic region:\[
\psi _{i,n}=\delta _{ji}\psi _{i,n}^{\textrm{incoming}}+\sum _{m=1}^{M_{j}}S_{mn}^{\left(ji\right)}\psi _{j,m}^{\textrm{outgoing}},\]
where $i$, $j$ specify leads, and $m$, $n$ specify channels, (iii)
express $S$-matrix in terms of $H_{in}$ and $N\times M$ matrix
$W$, which couples the subparts $S=I-2\pi iW^{\dagger }\left(E-H_{eff}\right)^{-1}W,$
with $H_{eff}=H_{in}-i\pi WW^{\dagger }$. This relation of $S$-matrix
to the effective Green's function $\left(E-H_{eff}\right)^{-1}$ is
the building block of the Hamiltonian approach to the system's statistics
in the universal (RMT) regime, i.e., the regime independent of the
details of the underlying classical dynamics \cite{Scattering}. The
non-Hermitian random matrices have been the subject of numerous works
on quantum chaotic scattering (see, for example, \cite{NonHermitianRM}
and references therein). 

Similar steps lead to the generalization of the effective non-Hermitian
Hamiltonian in the nonuniversal regime, as was done in a thick wire
in \cite{ThickWire} and for a disordered quantum dot, in which electrons,
confined by a hard wall potential can escape into leads, in \cite{footnote2}. 

In full analogy with the RMT problem one can take the same approach
to our problem involving open ballistic dot. For completeness, we
reproduce the main steps carried out in \cite{footnote2}, noting
that we deal with a two-dimensional clean electronic system.

In order to study the Green's functions $G$\begin{equation}
\left(E\pm i\epsilon -\mathcal{H}_{0}\right)G^{R,A}\left(\mathbf{r},\mathbf{r}'\right)=\delta \left(\mathbf{r}-\mathbf{r}'\right),\label{GoverningEq}\end{equation}
corresponding to the dot-plus-leads Hamiltonian\begin{equation}
\mathcal{H}_{0}=\left(\mathbf{p}-\frac{e}{c}\mathbf{A}\right)^{2}.\label{Eichzero}\end{equation}
 we introduce the crossectional surfaces $C_{n}$ perpendicular to
the walls of the leads close to the place of their attachment to the
dot. Then we reformulate the problem by specifying the boundary conditions
on these surfaces, thus eliminating the leads at the expense of modifying
the Hamiltonian of the system. First we introduce auxiliary retarded
and advanced Green's functions $\overline{G}^{R,A}\left(\mathbf{r},\mathbf{r}'\right)$
for each lead, satisfying Eq. (\ref{GoverningEq}) inside the region
of the lead bounded by the section $C_{n}$. The boundary condition
at $C_{n}$ is the only difference between $\overline{G}^{R,A}\left(\mathbf{r},\mathbf{r}'\right)$
and $G^{R,A}\left(\mathbf{r},\mathbf{r}'\right)$. The former vanishes
at this boundary. Considering $G^{R,A}\left(\mathbf{r},\mathbf{r}'\right)$,
such that $\mathbf{r}$ and $\mathbf{r}'$ belong to the regions on
the different sides from the section $C_{n}$ we make use of the following
identity (current conservation)\begin{equation}
\nabla _{r}\left[\left(\overline{G}^{A}\left(\mathbf{r},\mathbf{r}''\right)\right)^{*}\mathbf{v}_{r}G^{R}\left(\mathbf{r},\mathbf{r}'\right)+G^{R}\left(\mathbf{r},\mathbf{r}'\right)\left(\mathbf{v}_{r}\overline{G}^{A}\left(\mathbf{r},r''\right)\right)^{*}\right]=2i\delta \left(\mathbf{r}-\mathbf{r}''\right)G^{R}\left(\mathbf{r},\mathbf{r}'\right),\label{currentconservation}\end{equation}
where $\mathbf{v}_{r}=-i\left(\nabla _{r}+e\mathbf{A}/c\right)/m$.
To arrive at the current conservation relation (Eq. (\ref{currentconservation}))
we combine definitions of both $G^{R}$ and $\overline{G}^{A}$ (Eq.
(\ref{GoverningEq})), by multiplying them with $G^{R}\left(\mathbf{r},\mathbf{r}'\right)$
and subtracting from each other. Then, integration with respect to
$\mathbf{r}$-coordinates taken over the volume of the lead results
in\begin{equation}
G^{R}\left(\mathbf{r}'',\mathbf{r}'\right)=-\frac{i}{2}\int _{C_{n}}\left(\mathbf{x}_{n}\cdot \mathbf{v}_{r}\right)\left(\overline{G}^{A}\left(\mathbf{r},\mathbf{r}''\right)\right)^{*}G^{R}\left(\mathbf{r},\mathbf{r}'\right)d\mathbf{r},\label{CurrentConsIntegrated}\end{equation}
upon application of boundary condition on $\overline{G}^{A}\left(\mathbf{r},\mathbf{r}''\right)$.
Here $\mathbf{x}_{n}$ is the unit normal to $C_{n}$. Next, we apply
the velocity operator $\mathbf{v}_{r''}$ to both sides of Eq. (\ref{CurrentConsIntegrated})
and pick $\mathbf{r}''=\mathbf{y}$ on the crossection $C_{n}$ to
get\begin{equation}
\left(\mathbf{x}_{n}\cdot \mathbf{v}_{y}\right)G^{R}\left(\mathbf{y},\mathbf{r}'\right)=\int _{C_{n}}B_{n}\left(\mathbf{y},\mathbf{r}\right)G^{R}\left(\mathbf{r},\mathbf{r}'\right)d\mathbf{r},\label{NewBC}\end{equation}
where\[
B_{n}\left(\mathbf{y},\mathbf{r}\right)=\frac{i}{2}\left(\mathbf{x}_{n}\cdot \mathbf{v}_{y}\right)\left(\mathbf{x}_{n}\cdot \mathbf{v}_{r}\right)\overline{G}^{R}\left(\mathbf{y},\mathbf{r}\right),\]
and we used $\left(\overline{G}^{A}\left(\mathbf{r},\mathbf{r}''\right)\right)^{*}=\overline{G}^{R}\left(\mathbf{r}'',\mathbf{r}\right)$.
Note, that the functions $B_{n}\left(\mathbf{y},\mathbf{r}\right)$
are completely determined by the properties of the leads, which we
assumed to be ideal. Thus, the {}``dot-plus-leads'' system governed
by Eq. (\ref{GoverningEq}) is now reduced to the dot only, governed
by the same Eq. (\ref{GoverningEq}) and boundary conditions specified
by Eq. (\ref{NewBC}). We also note, that it is real part of a function
$B_{n}\left(\mathbf{y},\mathbf{r}\right)$ that is related to the
flux carried into the leads by the electronic states. To simplify
the boundary conditions we modify the Hamiltonian of the system even
further according to\begin{equation}
H=\mathcal{H}_{0}\mp \frac{i}{2}\sum _{n=1}^{N}\widehat{B}_{n}\delta C_{n},\label{effectiveHam}\end{equation}
where $\delta C_{n}$ is a surface $\delta $-function, defined via
$\int \delta C_{n}\Psi \left(\mathbf{r}\right)d\mathbf{r}=\int _{C_{n}}\Psi \left(\mathbf{r}\right)d\mathbf{r}$,
and $\widehat{B}_{n}\Psi \left(r\right)=\int _{C_{n}}\Re B_{n}\left(\mathbf{r},\mathbf{r}'\right)\, \Psi \left(\mathbf{r}'\right)d\mathbf{r}'$,
with $\Re B_{n}\left(\mathbf{r},\mathbf{r}'\right)=\gamma _{n}v_{n}\varphi _{n}\left(\mathbf{r}\right)\varphi _{n}\left(\mathbf{r}'\right)$.
To compacticify the formulas of the next sections we also assume that
each of the $N$ leads contains at most one open channel. Thus, we
end up looking for the Green's functions of the effective problem\begin{equation}
\left(E-H\right)G^{R,A}\left(\mathbf{r},\mathbf{r}'\right)=\delta \left(\mathbf{r}-\mathbf{r}'\right),\label{GoverningEq1}\end{equation}
associated with the Neumann boundary conditions\begin{equation}
\left(\mathbf{x}_{n}\cdot \mathbf{v}_{r}\right)\left.G^{R,A}\left(\mathbf{r},\mathbf{r}'\right)\right|_{C_{n}+0}=0,\label{bcNeumann}\end{equation}
where the derivatives are taken from the side of the lead (which is
indicated by $C_{n}+0$). The second term in Hamiltonian (Eq. (\ref{effectiveHam}))
describes the finite escape probability for the electrons colliding
with the boundary. Now we are ready to proceed with the construction
of the $\sigma $-model along the conventional lines developed for
closed systems \cite{Efetovsbook}. Note, that boundary condition
given by Eq. (\ref{bcNeumann}) ensures that system is effectively
closed, i.e. component of the current, which is normal to boundary,
vanishes.

\section{Supersymmetric generating functional}

Next we construct the generating functional for both retarded and
advanced Green's functions for Gaussian distributed dimensionless
coupling coefficients $\gamma _{n}$. These coefficients are related
to sticking probabilities and transmission coefficients, frequently
used within the Hamiltonian approach to chaotic scattering \cite{Scattering}
to compute statistical distributions of the resonance widths, delay
times and related characteristics. A detailed discussion of the physical
meaning of these coefficients can be found in \cite{Stockmannbook,Lobkis,LossySystems,Mine}
and in references therein. In order to perform non-perturbative calculations
of these averages in the non-universal regime we first construct the
supersymmetric functional $Z$ \cite{Efetovsbook,Mirlin}. Any correlator
of the dot Green's functions can be later obtained from it by differentiation
with respect to sources $J$ \cite{WhiteBook} as long as we know
the average $Z\left[J\right]$ over the $\gamma $-ensemble.

We can assume the magnetic field (Eq. (\ref{Eichzero})) to be vanishingly
small and remove it from consideration. Its role is reduced to breaking
the time reversal symmetry, and justifying the use $4$-component
supervectors $\Psi \left(\mathbf{r}\right)^{T}=\left\{ S_{1}\left(\mathbf{r}\right),\chi _{1}\left(\mathbf{r}\right),S_{2}\left(\mathbf{r}\right),\chi _{2}\left(\mathbf{r}\right)\right\} $
in the supersymmetric functional. The generalization of our formalism
to other symmetry classes is quite straightforward and will require
doubling the space \cite{Efetovsbook}. It is convenient to express
the coupling coefficients as a sum of constant and stochastic parts:
$\gamma _{n}=\hat{\gamma }_{n}+\widetilde{\gamma }_{n}$. For the
statistics of $\widetilde{\gamma }_{n}$ we assume that $\left\langle \widetilde{\gamma }_{n}\right\rangle =0,$
$\left\langle \widetilde{\gamma }_{n}\widetilde{\gamma }_{m}\right\rangle =x_{n}^{2}\delta _{nm}$;
all higher moments factorize into second moments. We indicate averaging
over random couplings to the leads by the shorthand notation $\left\langle \dots \right\rangle _{\widetilde{\gamma }}$.
Then we {}``eliminate the leads'' \cite{footnote2}, passing to
the Hamiltonian given by Eq. (\ref{effectiveHam}). 

In terms of supervectors $\Psi \left(\mathbf{r}\right)$, and supermatrices
$L=\textrm{diag}\left\{ 1,1-1,1\right\} $, $\Lambda =\textrm{diag}\left\{ 1,1-1,-1\right\} $
\cite{Mirlin} the generating functional $Z\left[J\right]$ is written
down as follows\begin{equation}
\left\langle Z\left[J\right]\right\rangle _{\gamma }=\int d\Psi ^{*}d\Psi \textrm{e}^{-\mathcal{L}\left[\Psi \right]}\left\langle \textrm{e}^{-\mathcal{L}_{\delta }\left[\Psi \right]}\right\rangle _{\tilde{\gamma }},\label{generatingunctional}\end{equation}
\[
\mathcal{L}\left[\Psi \right]=i\int \Psi ^{\dagger }\left(\mathbf{r}\right)\widehat{\mathcal{H}}_{J}L\, \Psi \left(\mathbf{r}\right)d\mathbf{r}+\frac{1}{2}\sum _{n=1}^{N}v_{n}\hat{\gamma }_{n}\int _{C_{n}}\Psi ^{\dagger }\left(y_{n}\right)\varphi _{n}\left(y_{n}\right)\varphi _{n}\left(y'_{n}\right)\Lambda L\, \Psi \left(y'_{n}\right),\]
\[
\mathcal{L}_{\delta }\left[\Psi \right]=\sum _{n=1}^{N}\frac{\widetilde{\gamma }_{n}v_{n}}{2}\int _{C_{n}}\Psi ^{\dagger }\left(y_{n}\right)\varphi _{n}\left(y_{n}\right)\varphi _{n}\left(y'_{n}\right)\Lambda L\, \Psi \left(y'_{n}\right),\]
with\[
\widehat{\mathcal{H}}_{J}=\left(-\frac{\nabla ^{2}}{2m}-E\right)I_{4}+i\epsilon \Lambda +J,\]
where $\epsilon $ is infinitesimally small, $\varphi _{n}\left(y\right)=\sqrt{2/d_{n}}\sin \left(\pi y/d_{n}\right)$
(for hard-wall lead of width $d_{n}$). Here $\int _{C_{n}}$ stands
for a double integration over $y_{n}$ and $y'_{n}$, the transverse
coordinates along the crossection $C_{n}$. The exact form of the
source supermatrix\[
J=\textrm{diag}\left\{ J_{1}\left(\mathbf{r}\right),J_{2}\left(\mathbf{r}\right),J_{1}\left(\mathbf{r}\right),J_{2}\left(\mathbf{r}\right)\right\} ,\]
is dictated by the choice of the physical quantity we eventually wish
to calculate.

Averaging over $\widetilde{\gamma }_{n}$ produces:\begin{equation}
\left\langle \textrm{e}^{-\mathcal{L}_{\delta }\left[\Psi \right]}\right\rangle _{\gamma }=\left\langle 1+\frac{1}{8}\sum _{n=1}^{N}x_{n}^{2}v_{n}^{2}\left\{ \int _{C_{n}}\Psi ^{\dagger }\left(y_{n}\right)L\varphi _{n}\left(y_{n}\right)\varphi _{n}^{*}\left(y'_{n}\right)\Psi \left(y'_{n}\right)\right\} ^{2}+\dots \right\rangle _{\tilde{\gamma }}.\label{AveragedLd}\end{equation}
This step towards constructing supersymmetric NL$\sigma $M for our
system is no more difficult than in the problems requiring averaging
over impurities or random magnetic field. As we shall see below it
results in the coupling of the effective field with the source of
boundary transmission.

\section{Effective problem for a smooth convex boundary}

Next we carry out the Hubbard-Stratonovich transformation to decouple
the {}``interaction'' terms (Eq. (\ref{AveragedLd})) which entered
the action of Eq. (\ref{generatingunctional}) after averaging. The
procedure, involving supermatrix fields $Q_{n}$, is explained in
Appendix A and leads to\begin{align}
\left\langle Z\left[J\right]\right\rangle _{\gamma } & =\int \prod _{n}DQ_{n}\int d\Psi ^{*}d\Psi \textrm{e}^{-\mathcal{L}\left[\Psi \right]}\exp \sum _{n=1}^{N}\left\{ \frac{x_{n}v_{n}m}{2}\int _{C_{n}}\Psi ^{\dagger }\left(y_{n}\right)L\right.\nonumber \\
 & \times \varphi _{n}\left(y_{n}\right)\varphi _{n}^{*}\left(y'_{n}\right)\Psi \left(y'_{n}\right)\int _{C_{n}}Q_{n}\left(y''_{n},y'''_{n}\right)\varphi _{n}\left(y''_{n}\right)\varphi _{n}^{*}\left(y'''_{n}\right)\nonumber \\
 & \left.-\frac{m^{2}}{2}\left(\int _{C_{n}}Q_{n}\left(y_{n},y'_{n}\right)\varphi _{n}\left(y_{n}\right)\varphi _{n}^{*}\left(y'_{n}\right)\right)^{2}\right\} .\label{Qintroduced}
\end{align}
 Next, we make further simplifications by setting $d_{n}=d$ and distributing
the large number of narrow leads $d_{n}=d\ll P$ (where $P$ is perimeter
of the dot), close to each other all around the boundary. In this
limit we can approximately write $\varphi _{n}\left(y_{n}\right)\varphi _{n}^{*}\left(y'_{n}\right)\simeq 2/d$.
These two model assumptions are not essential for subsequent analysis,
although they do make it more transparent. The distribution of the
leads coupling coefficients can be chosen piece-wise continuous, while
widths can be given as a set of input parameters to the problem at
hand together with $\hat{\gamma }_{n}$, $x_{n}$ and $v_{n}$.

Then we make use of the mean value theorem for each of the $\int _{C_{n}}$
integrals, provided all the integrands are smooth functions of $y_{n}$.
The resulting sum in the exponent of Eq. (\ref{Qintroduced}) can
be combined into a single integral over the arclength $s$, ranging
from $0$ to $P$, which runs along the boundary, defined in polar
coordinates as $r=R\left(s\right)$, producing\begin{align*}
\left\langle Z\left[J\right]\right\rangle _{\gamma } & =\int DQ\left(s\right)\int d\Psi ^{*}\left(r,\theta \right)d\Psi \left(r,\theta \right)\textrm{e}^{-\mathcal{L}\left[\Psi \right]}\\
\times \exp  & \left\{ 2md\int _{0}^{P}\widetilde{x}\left(s\right)\Psi ^{\dagger }\left(R\left(s\right),s\right)LQ\left(s\right)\Psi \left(R\left(s\right),s\right)ds-2m^{2}d\int _{0}^{P}Q^{2}\left(s\right)ds\right\} ,
\end{align*}
where $\widetilde{x}\left(s\right)=x\left(s\right)v\left(s\right)$
and both band velocity $v\left(s\right)$, mean $\hat{\gamma }\left(s\right)$
and r.m.s. $x\left(s\right)$ of the coupling strength are now smooth
functions of arclength. The corresponding changes in the term $\mathcal{L}\left[\Psi \right]$
are made accordingly. Upon carrying out the $\Psi $-integration,
we arrive at the following representation of the generating functional
in polar coordinates $\mathbf{r}=\left(r,\theta \right)$\begin{align}
\left\langle Z\left[J\right]\right\rangle _{\gamma } & =\int DQ\exp \, \textrm{Str}\left[-2m^{2}d\int d\mathbf{r}d\mathbf{r}'Q^{2}\left(s\right)\delta _{P}\delta \left(\mathbf{r}-\mathbf{r}'\right)\right.\nonumber \\
 & -\ln \left\{ \left(-i\widehat{\mathcal{H}}_{0}+\frac{v\left(s\right)\hat{\gamma }\left(s\right)}{d}\Lambda \delta _{P}\right)\delta \left(\mathbf{r}-\mathbf{r}'\right)-\frac{md\widetilde{x}\left(s\right)}{2}Q\left(s\right)\delta _{P}\delta \left(\mathbf{r}-\mathbf{r}'\right)\right\} \nonumber \\
 & \left.-\ln \left(I_{4}+\epsilon \Lambda \mathcal{G}\left(\mathbf{r},\mathbf{r}'\right)+\mathcal{G}\left(\mathbf{r},\mathbf{r}'\right)J\left(\mathbf{r}\right)\right)\right],\label{FreeEnergyInitial}
\end{align}
where $\delta _{P}$ is a perimeter delta-function and the supermatrix
{}``Green's function'' $\mathcal{G}$ is determined from\begin{equation}
\left\{ \widehat{\mathcal{H}}_{0}-i2m\widetilde{x}\left(s\right)d\left(Q\left(s\right)-\frac{\hat{\gamma }\left(s\right)}{2x\left(s\right)md}\Lambda \right)\delta _{P}\right\} \mathcal{G}\left(\mathbf{r},\mathbf{r}'\right)=i\delta \left(\mathbf{r}-\mathbf{r}'\right),\label{Greensfunction}\end{equation}
and $\widehat{\mathcal{H}}_{0}=\widehat{\mathcal{H}}_{J}\left(J=0\right).$
Thus we reduced the generating functional for our system to the integral
over the supermatrix superfield with boundary support. We have\begin{equation}
\left\langle Z\left[J\right]\right\rangle _{\gamma }=\int DQ\textrm{e}^{F\left[Q\right]+F_{J}\left[\mathcal{G}\right]},\label{ZJintermsofQ}\end{equation}
with the free energy\[
F\left[Q\right]=\textrm{Str}\int d\mathbf{r}d\mathbf{r}'\left\{ -2m^{2}dQ\left(s\right)^{2}\delta _{P}\delta \left(\mathbf{r}-\mathbf{r}'\right)+\ln -i\mathcal{G}^{-1}\left(\mathbf{r},\mathbf{r}'\right)\right\} ,\]
and symmetry breaking terms\[
F_{J}\left[\mathcal{G}\right]=-\textrm{Str}\int \ln \left(I_{4}+\epsilon \Lambda \mathcal{G}\left(\mathbf{r},\mathbf{r}'\right)+\mathcal{G}\left(\mathbf{r},\mathbf{r}'\right)J\left(\mathbf{r}\right)\right)d\mathbf{r}d\mathbf{r}'.\]
In order to reduce our generating functional (Eq. (\ref{ZJintermsofQ}))
to a NL$\sigma $M we employ the saddle point condition, which in
our case reads\begin{equation}
Q_{sp}\left(s\right)=\frac{\widetilde{x}\left(s\right)}{2m}\mathcal{G}\left(R\left(s\right),R\left(s\right),s,s,Q_{sp}\left(s\right)\right).\label{SaddlePC}\end{equation}
We assume the saddle point solution to be diagonal: $Q_{sp}\left(s\right)=Q_{0}\left(s\right)\Lambda $.
To proceed with the analysis of fluctuations, one needs to determine
both the $Q_{0}\left(s\right)$ and the diagonal Green's function
supermatrix $\mathcal{G}_{sp}\left(\mathbf{r},\mathbf{r}'\right)$.
Thus, by combining Eqs. (\ref{Greensfunction}) and (\ref{SaddlePC})
with the assumption about saddle point structure we mapped the original
problem with random boundary condition onto an effective problem specified
by the differential equation:\begin{equation}
\frac{1}{2m}\left(\nabla ^{2}-\kappa ^{2}\right)\mathcal{G}\left(r,r',\theta ,\theta ',\kappa ^{2}\right)=-\frac{i\delta \left(r-r'\right)\delta \left(\theta -\theta '\right)}{r},\label{ModifiedHelmholtz}\end{equation}
where $\kappa ^{2}=-2mE$, with associated boundary conditions\begin{equation}
\frac{\partial }{\partial r}\mathcal{G}\left.\left(r,r',\theta ,\theta ',\kappa ^{2}\right)\right|_{S^{-}}=i\frac{f\left(s\right)}{R\left(s\right)}\mathcal{G}\left.\left(r,r',\theta ,\theta ',\kappa ^{2}\right)\right|_{S^{-}},\label{BC}\end{equation}
\begin{equation}
\frac{\partial }{\partial r}\mathcal{G}\left.\left(r,r',\theta ,\theta ',\kappa ^{2}\right)\right|_{S^{+}}=0,\label{BCzero}\end{equation}
where $f\left(s\right)=m^{2}\widetilde{x}\left(s\right)\widetilde{Q}_{0}\left(s\right)R\left(s\right)d$,
$S^{-}$, $S^{+}$ are the inner and the outer surfaces of the dot,
and $\widetilde{Q}_{0}\left(s\right)=Q_{0}\left(s\right)-\hat{\gamma }\left(s\right)/\left(2x\left(s\right)md\right)$.

\section{Born-Oppenheimer-like approximation}

To construct the Greens function of Eq. (\ref{ModifiedHelmholtz})
we employ the technique of \cite{Stewartson}, which was also used
in \cite{arxivePaper} for a circular billiard. In the latter case
the corresponding Green's function reads

\begin{equation}
\mathcal{G}_{\textrm{circle}}\left(r,r',\theta ,\theta ',\kappa ^{2}\right)=\frac{im}{\pi }\sum _{n=-\infty }^{\infty }I_{n}\left(\kappa r_{<}\right)\left\{ a_{n}I_{n}\left(\kappa r_{>}\right)+K_{n}\left(\kappa r_{>}\right)\right\} \textrm{e}^{in\left(\theta -\theta '\right)},\label{GreensfunctionIK}\end{equation}
with $r_{>}$ ($r_{<}$) is a maximum (minimum) of $r$ and $r'$,
$I_{n}$ and $K_{n}$ are modified Bessel functions respectively,
and the coefficients $a_{n}$ are chosen to ensure the boundary condition.
The summation in Eq. (\ref{GreensfunctionIK}) is replaced with integration,
while Bessel functions are replaced with their uniform approximations
\cite{Stewartson}. 

For a generic convex smooth-wall billiard a similar expansion is possible
with the help of the recently proposed Born-Oppenheimer-like approximation
\cite{Zaitsev1}. The method used in \cite{Zaitsev1} assumes that
the angular variation of $x\left(s\right)$, $f\left(s\right)$, etc.
is slow and enables one to determine certain classes of eigenstates
quite accurately. Guided by the example of circular billiard, and
the fact that WGM have the longest life-time, we focus our attention
on the limit of large angular momentum. 

The Greens function (Eq. (\ref{GreensfunctionIK})) was build from
two radial solutions of Eq. (\ref{ModifiedHelmholtz}) and angular
harmonics. In case of a generic billiard, the angular {}``part''
of the eigenstate is obtained from WKB solution of slow-variable equation
\cite{Zaitsev1},\[
\psi \left(s\right)=\exp \left\{ i\lambda \int ^{s}\frac{F\left(\overline{s}\right)}{R\left(\overline{s}\right)}d\overline{s}\right\} \]
while the radial {}``part'' is given by the linear combination of
$I_{l\left(s\right)}\left(\kappa r_{s}\right)$ and $K_{l\left(s\right)}\left(\kappa r_{s}\right)$.
Here angular momentum {}``index'' is parametrized as $l\left(s\right)=\lambda F\left(s\right)$,
$r_{s}$ is a local radial coordinate, computed from the local center
of curvature \cite{Zaitsev1}. Eigenvalues $\lambda $ are determined
from arclength quantization condition ($\psi \left(0\right)=\psi \left(P\right)$):
$\lambda _{n}=2\pi n/\overline{\left(\frac{F\left(s\right)}{R\left(s\right)}\right)}P$.
Here and below we use {}``overbar'' to denote the perimeter average:
$\overline{R}=\int _{0}^{P}R\left(s\right)ds/P$, $\overline{f}=\int _{0}^{P}f\left(s\right)ds/P$,
etc. Note that for dimensionless quantities we will not use overall
factor $1/P$. The explicit form of $F\left(s\right)$ can be found
from the radial quantization condition \cite{Zaitsev1}. Below we
consider $F\left(s\right)$ to be known. Thus, the Green's function
for the effective problem given by Eqs. (\ref{ModifiedHelmholtz},\ref{BC})
can be approximated as\begin{align}
\mathcal{G}\left(r,r',\theta ,\theta ',\kappa ^{2}\right) & =\frac{imR\left(s\right)}{P}\sum _{n=-\infty }^{\infty }I_{\lambda _{n}F\left(s\right)}\left(\kappa r_{s<}\right)\left\{ a_{n}I_{\lambda _{n}F\left(s\right)}\left(\kappa r_{s>}\right)+K_{\lambda _{n}F\left(s\right)}\left(\kappa r_{s>}\right)\right\} \nonumber \\
 & \times \exp \left\{ i\lambda _{n}\int _{s'}^{s}\frac{F\left(\overline{s}\right)}{R\left(\overline{s}\right)}d\overline{s}\right\} ,\label{GreensFunctionIK1}
\end{align}
with\[
a_{n}\left(s\right)=\frac{-if\left(s\right)K_{\lambda _{n}F\left(s\right)}\left(\kappa R\left(s\right)\right)+\kappa R\left(s\right)K'_{\lambda _{n}F\left(s\right)}\left(\kappa R\left(s\right)\right)}{if\left(s\right)I_{\lambda _{n}F\left(s\right)}\left(\kappa R\left(s\right)\right)-\kappa R\left(s\right)I'_{\lambda _{n}F\left(s\right)}\left(\kappa R\left(s\right)\right)}.\]
This expression does not hold far from the boundary, but it suits
our purposes. For example, it can be used to determine saddle point
solution $Q_{0}\left(s\right)$. We have, at the billiard boundary\begin{align*}
\mathcal{G}\left(R\left(s\right),R\left(s\right),s,s',\kappa ^{2}\right) & =\frac{imR\left(s\right)}{P}\sum _{n=-\infty }^{\infty }\frac{I_{\lambda _{n}F\left(s\right)}\left(\kappa R\left(s\right)\right)}{if\left(s\right)I_{\lambda _{n}F\left(s\right)}\left(\kappa R\left(s\right)\right)-\kappa R\left(s\right)I'_{\lambda _{n}F\left(s\right)}\left(\kappa R\left(s\right)\right)}\\
 & \times \exp \left\{ i\lambda _{n}\int _{s'}^{s}\frac{F\left(\overline{s}\right)}{R\left(\overline{s}\right)}d\overline{s}\right\} ,
\end{align*}
Then, $Q_{0}\left(s\right)$ is determined by the stationary point
condition (Eq. (\ref{SaddlePC}))\begin{align}
i\frac{\widetilde{x}\left(s\right)R\left(s\right)}{2P} & \sum _{n=-\infty }^{\infty }\frac{I_{\lambda _{n}F\left(s\right)}\left(\kappa R\left(s\right)\right)}{if\left(s\right)I_{\lambda _{n}F\left(s\right)}\left(\kappa R\left(s\right)\right)-\kappa R\left(s\right)I'_{\lambda _{n}F\left(s\right)}\left(\kappa R\left(s\right)\right)}\nonumber \\
 & =\widetilde{Q}_{0}\left(s\right)+\frac{\hat{\gamma }\left(s\right)}{2x\left(s\right)md}.\label{SumIdentity}
\end{align}
Dropping the imaginary part of $\widetilde{Q}_{0}\left(s\right)$,
we set $\widetilde{g}\left(s\right)=\kappa R\left(s\right)$, and
proceed as follows. We evaluate the sum over $n$ in Eq. (\ref{SumIdentity})
asymptotically in the limit: $\widetilde{g}\left(s\right)\gg 1$,
$f\left(s\right)/\widetilde{g}\left(s\right)\sim 1$ by replacing
it with integral over corresponding continuous variable. Following
the technique, described in Ref. \cite{Stewartson}, and illustrated
in Appendix B for a more involved calculation, we use uniform approximation
for the Bessel function and its derivative \cite{Abramowitz}, carry
out the integration and rewrite Eq. (\ref{SumIdentity}) as \[
\frac{\widetilde{x}\left(s\right)R\left(s\right)\overline{\left(F\left(s\right)/R\left(s\right)\right)}}{2F\left(s\right)\sqrt{f\left(s\right)^{2}-g\left(s\right)^{2}}}=\widetilde{Q}_{0}\left(s\right)+\frac{\hat{\gamma }\left(s\right)}{2x\left(s\right)md}.\]
 after the substitution $\widetilde{g}\left(s\right)\rightarrow -ig\left(s\right)$
($\kappa \rightarrow -ik$), to the leading order in $1/g\left(s\right)$.
This equation can be now solved numerically yielding $f\left(s\right)$
(or $\widetilde{Q}_{0}\left(s\right)$) for any prescribed set of
parameters of the problem.

\section{Ballistic nonlinear $\sigma $-model}

Having determined the relation between the Green's function $\mathcal{G}$
and saddle-point value of superfield $Q$, we come back to further
analyze the generating functional specified by Eq. (\ref{ZJintermsofQ}).
As a final step in derivation of NL$\sigma $M we have to expand the
action $F\left[Q\right]$ up to quadratic order around the extremum
and include first variation of symmetry breaking terms in $F_{J}\left[\mathcal{G}\right]$.
Observing that $F\left[Q_{sp}\right]=0$ we turn to the fluctuations
of $Q$, which decompose into a transverse piece $\delta Q^{(t)}$
(along the saddle-point manifold \cite{Efetovsbook}) and a longitudinal
piece $\delta Q^{(l)}$ (orthogonal to the saddle point manifold).
We focus on the transverse part of the action, the part anticommuting
with the $\Lambda $-like saddle-point solution. Our goal is to demonstrate
that in absence of symmetry breaking terms transverse fluctuations
are massless (Goldstone) modes of the theory. The purely transverse
terms are given by \cite{Efetovsbook}\begin{align*}
F_{t}\left[\delta Q\right] & =-4m^{2}d\, \textrm{Str}\int _{0}^{P}\left(\delta Q^{\left(t\right)}\left(s\right)\right)^{2}ds+\left(md\right)^{2}\textrm{Str}\int _{0}^{P}\int _{0}^{P}\mathcal{G}\left(Q_{0}\left(s\right)\right)\mathcal{G}\left(-Q_{0}\left(s\right)\right)\\
 & \times \widetilde{x}\left(s\right)\widetilde{x}\left(s'\right)\delta Q^{\left(t\right)}\left(s\right)\delta Q^{\left(t\right)}\left(s'\right)dsds'.
\end{align*}
In view of the developments of the previous Section, we can expand
deviations of $Q$ in approximate angular eigenstates: $\delta Q^{\left(t\right)}\left(s\right)=\sum _{l=-\infty }^{\infty }Q_{l}^{\left(t\right)}\exp \left\{ i\lambda _{l}\int ^{s}d\overline{s}F\left(\overline{s}/R\left(\overline{s}\right)\right)\right\} $,
and, setting $\mathcal{G}\left(Q_{0}\right)=\mathcal{G}$, and $\mathcal{G}\left(-Q_{0}\right)=\widetilde{\mathcal{G}}$
we obtain\begin{align}
F_{t}\left[\delta Q\right] & =-4m^{2}d\, \textrm{Str}\sum _{l,n}\int _{0}^{P}\delta Q_{l}^{\left(t\right)}\delta Q_{n}^{\left(t\right)}\textrm{e}^{i\left(\lambda _{l}+\lambda _{n}\right)\int ^{s}\frac{F\left(\overline{s}\right)}{R\left(\overline{s}\right)}d\overline{s}}ds+\left(md\right)^{2}\textrm{Str}\sum _{l,n}\delta Q_{l}^{\left(t\right)}\delta Q_{n}^{\left(t\right)}\nonumber \\
 & \times \int _{0}^{P}ds\int _{0}^{P}ds'\mathcal{G}\widetilde{\mathcal{G}}\widetilde{x}\left(s\right)\widetilde{x}\left(s'\right)\textrm{e}^{i\lambda _{l}\int ^{s}\frac{F\left(\overline{s}\right)}{R\left(\overline{s}\right)}d\overline{s}+i\lambda _{n}\int ^{s'}\frac{F\left(\overline{s}\right)}{R\left(\overline{s}\right)}d\overline{s}}.\label{TransvAction}
\end{align}
Our intermediate goal now is to get a Ward-like identity, which would
allow us to evaluate massive part of the transverse action. At this
point we consider Eq. (\ref{Greensfunction}) for both Green's functions,\begin{equation}
\left(\frac{\nabla ^{2}}{2m}-i\frac{f\left(s\right)}{2mR\left(s\right)}\delta _{P}-\frac{\kappa ^{2}}{2m}\right)\mathcal{G}=\frac{i}{r}\delta \left(r-r'\right)\delta \left(\theta -\theta '\right),\label{EqnforG}\end{equation}
\begin{equation}
\left(\frac{\nabla ^{2}}{2m}+i\frac{f\left(s\right)}{2mR\left(s\right)}\delta _{P}-\frac{\kappa ^{2}}{2m}\right)\widetilde{\mathcal{G}}=\frac{i}{r}\delta \left(r-r'\right)\delta \left(\theta -\theta '\right),\label{EqnforGtilde}\end{equation}
 and multiply Eq. (\ref{EqnforG}) by $\widetilde{\mathcal{G}}h\left(s,s'\right)$
and Eq. (\ref{EqnforGtilde}) with $\mathcal{G}h\left(s,s'\right)$.
We are free to chose function $h\left(s,s'\right)$ to behave as we
like inside the billiard, because only the value of this function
at the boundary matters for manipulations in Eq. (\ref{TransvAction}).
In particular we may require that $h$ slowly decay to zero in the
radial variable as we move away from the boundary. Since the radial
rate of variation is approximately $1/\overline{R}$, while the angular
rate is $1/\lambda _{F}$ (where $\lambda _{F}$ is a Fermi wavelength),
we neglect all radial derivatives in the subsequent analysis, which
is correct to leading order in $\overline{R}/\lambda _{F}$. Next,
we subtract one equation from the other, integrate the resulting expression
over the area of the dot $D$ (including the boundary: $0<r<R^{+}\left(s\right)$),
and set the remaining free radial coordinate to $R\left(s\right)$
to get\begin{align}
\int _{V}\left(h\widetilde{\mathcal{G}}\nabla \cdot \nabla \mathcal{G}-h\mathcal{G}\nabla \cdot \nabla \widetilde{\mathcal{G}}\right)d\tau  & =2i\int _{0}^{P}dsf\left(s\right)h\left(s,s'\right)\nonumber \\
\times \mathcal{G}\left(R\left(s\right),R\left(s'\right),s,s'\right) & \widetilde{\mathcal{G}}\left(R\left(s\right),R\left(s'\right),s,s'\right)+2imh\left(s',s'\right)\nonumber \\
\times \left(\widetilde{\mathcal{G}}\left(R\left(s'\right),R\left(s'\right),s',s'\right)\right. & \left.-\mathcal{G}\left(R\left(s'\right),R\left(s'\right),s',s'\right)\right).\label{Identity1}
\end{align}
Note, that according to the saddle-point condition (Eq. (\ref{SaddlePC}))\begin{equation}
\widetilde{\mathcal{G}}\left(R\left(s\right),R\left(s\right),s,s\right)-\mathcal{G}\left(R\left(s\right),R\left(s\right),s,s\right)=-2m\frac{\widetilde{Q}_{0}\left(s\right)}{\widetilde{x}\left(s\right)},\label{SPidentity}\end{equation}
 which might be used in the last line in Eq. (\ref{Identity1}). Furthermore,
one can transform the left-hand side of the Eq. (\ref{Identity1})
by making use of Gauss's theorem for the domain $D$ (with boundary
$\partial D$) \begin{align}
\int _{D}\left(h\widetilde{\mathcal{G}}\nabla \cdot \nabla \mathcal{G}-h\mathcal{G}\nabla \cdot \nabla \widetilde{\mathcal{G}}\right)d\tau  & =\int _{\partial D}h\left(\widetilde{\mathcal{G}}\nabla \mathcal{G}-\mathcal{G}\nabla \widetilde{\mathcal{G}}\right)\cdot d\sigma \nonumber \\
-\int _{D} & \left(\nabla \left(h\widetilde{\mathcal{G}}\right)\cdot \nabla \mathcal{G}-\nabla \left(h\mathcal{G}\right)\cdot \nabla \widetilde{\mathcal{G}}\right)d\tau .\label{Identity2}
\end{align}
Application of the boundary conditions (Eq. (\ref{BCzero})) to Eq.
(\ref{Identity2}) makes the surface integral vanish. It remains to
evaluate the area integral on the right-hand side of Eq. (\ref{Identity2}),
which we now substitute into Eq. (\ref{Identity1}) and obtain, neglecting
the terms with radial derivatives of $h$\begin{align}
\int _{0}^{2\pi }\frac{\partial h\left(s,s'\right)}{\partial \theta }d\theta \int _{0}^{R\left(s\right)}\frac{dr}{r}\left(\widetilde{\mathcal{G}}\frac{\partial }{\partial \theta }\mathcal{G}-\mathcal{G}\frac{\partial }{\partial \theta }\widetilde{\mathcal{G}}\right) & +4im^{2}h\left(s',s'\right)\frac{\widetilde{Q}_{0}\left(s\right)}{\widetilde{x}\left(s\right)}\nonumber \\
 & =2i\int _{0}^{P}dsh\left(s,s'\right)\widetilde{\mathcal{G}}\mathcal{G},\label{2ndterminaction}
\end{align}
where we used Eq. (\ref{SPidentity}). The boundary value of function
$h$ is completely at our discretion. We select\[
h\left(s,s'\right)=-i\left(md\right)^{2}\sum _{l,n}\frac{\widetilde{x}\left(s\right)\widetilde{x}\left(s'\right)}{f\left(s\right)}\textrm{e}^{i\lambda _{l}\int ^{s}\frac{F\left(\overline{s}\right)}{R\left(\overline{s}\right)}d\overline{s}+i\lambda _{n}\int ^{s'}\frac{F\left(\overline{s}\right)}{R\left(\overline{s}\right)}d\overline{s}}\delta Q_{l}^{\left(t\right)}\delta Q_{n}^{\left(t\right)}.\]
Then we integrate both sides of Eq. (\ref{2ndterminaction}) with
respect to $s'$ along the boundary of the billiard and substitute
the result together with $h$ in each term of the sum in the second
term of $F_{t}\left[\delta Q\right]$ (Eq. (\ref{TransvAction})).

We observe that the massive contribution coming from variation of
$Q\left(s\right)^{2}$ (the first term on the right-hand side of Eq.
(\ref{TransvAction})) will be canceled by the second term on the
left hand side of Eq. (\ref{2ndterminaction}). The surviving linear
in $\delta Q^{\left(t\right)}$ term proportional to $\hat{\gamma }\left(s\right)$
(cf. \cite{arxivePaper}) will be accounted later, together with the
symmetry breaking terms. In addition to that, we need to demonstrate
that massive longitudinal modes $\delta Q^{\left(l\right)}$ are decoupled
from the essentially massless transverse modes. Then, the longitudinal
fluctuations around saddle-point can be all integrated out, producing
unity due to supersymmetry.

To complete the derivation of NL$\sigma $M we still have to analyze
the remaining part of the transverse action\begin{equation}
F_{t}\left[\delta Q\right]=\textrm{Str}\sum _{l,n}\int _{0}^{P}ds'\int _{0}^{P}ds\int _{0}^{R\left(s\right)}\frac{\partial h\left(s,s'\right)}{\partial s}R^{2}\left(s\right)\left(\widetilde{\mathcal{G}}\frac{\partial }{\partial s}\mathcal{G}-\mathcal{G}\frac{\partial }{\partial s}\widetilde{\mathcal{G}}\right)\frac{dr}{r}\label{Diffusionaction}\end{equation}
 We perform this step using the asymptotic technique, we employed
in previous Section. After a straightforward, but lengthy procedure,
which we allocate to the Appendix B we obtain\begin{equation}
F_{t}\left[\delta Q\right]\simeq -\textrm{Str}D_{0}\int _{0}^{P}\left(\frac{R\left(s\right)}{F\left(s\right)}\frac{\partial \delta Q^{\left(t\right)}}{\partial s}\right)^{2}ds,\label{Diffusivepart}\end{equation}
 \[
D_{0}=\overline{\left(\frac{F}{R}\right)}\frac{m^{4}d^{2}\overline{\tilde{x}}}{4P}\overline{\left(\frac{\tilde{x}FR^{2}g\left(2g^{2}+f^{2}\right)}{f^{4}\sqrt{f^{2}-g^{2}}}\right)}.\]
 One can use similar set of manipulations for the purely longitudinal
and proportional to $\delta Q^{\left(l\right)}\delta Q^{\left(t\right)}+\delta Q^{\left(t\right)}\delta Q^{\left(l\right)}$
parts of the action, which consist of the fluctuations commuting with
the saddle-point solution, and therefore contain $\mathcal{G}^{2}$
in the contribution to second variation coming from the logarithmic
term in $F\left[Q\right]$ (cf. Eq. (\ref{TransvAction})). These
manipulations enable us to show that to the considered order in $1/\overline{g}$
the longitudinal and transverse modes are decoupled (the coefficient
in front of the $\delta Q^{\left(l\right)}\delta Q^{\left(t\right)}+\delta Q^{\left(t\right)}\delta Q^{\left(l\right)}$
is small compared to the mass of the longitudinal modes). 

To finish the construction of the nonlinear $\sigma $-model, we integrate
out the longitudinal modes, set $\delta Q^{\left(t\right)}=Q$, and
expand the symmetry breaking terms $F_{J}\left[\mathcal{G}\left(Q\right)\right]$
to the lowest order in $J$ and $\epsilon $ to get\[
\left\langle Z\left[J\right]\right\rangle _{\gamma }=\int DQ\textrm{e}^{-F\left[Q\right]},\]
with the free energy given by\begin{align}
F\left[Q\right] & =\textrm{Str}\int d\mathbf{r}d\mathbf{r}'\left[\left\{ D_{0}\left(\frac{R}{F}\frac{\partial Q\left(s\right)}{\partial s}\right)^{2}+\frac{\overline{\gamma }}{xmd}Q\left(s\right)\Lambda \right\} \delta _{P}\delta \left(\mathbf{r}-\mathbf{r}'\right)\right.\nonumber \\
 & \left.+\int ds''\left(\epsilon \Lambda +J\left(\mathbf{r}\right)\right)Q\left(s''\right)a\left(\mathbf{r},R\left(s''\right)s'';\mathbf{r}',R\left(s''\right)s''\right)\right]\label{FreeEnergyfinal}
\end{align}
where\[
a\left(\mathbf{r},R\left(s''\right)s'';\mathbf{r}',R\left(s''\right)s''\right)=i\frac{md\tilde{x}}{2}\mathcal{G}_{sp}\left(R\left(s''\right)s'',\mathbf{r}'\right)\mathcal{G}_{sp}\left(\mathbf{r},R\left(s''\right)s''\right)\]
and we suppressed explicit slow $s$-dependence everywhere. The free
energy given by Eq. (\ref{FreeEnergyfinal}) is the central result
of our paper. It displays the surface modes $Q\left(s\right)$, which
undergo diffusion and drift, and are coupled to the interior of the
dot by the last term. 

For the calculation of any physical quantity, expressed via Green's
function correlators such as, e.g. $\left\langle G^{R}\left(\mathbf{r,r}'\right)G^{A}\left(\mathbf{r,r}'\right)\right\rangle _{\gamma }$,
it is necessary to have actual parametrization for $Q\left(s\right)$.
Since basic symmetries of our problem are not any different from these
of the corresponding $Q\left(\mathbf{r}\right)$-field of the diffusive
problem \cite{WhiteBook,Efetovsbook,Mirlin} our supermatrix $Q$
can be parametrized as suggested in \cite{Efetovsbook,Mirlin}.

\section{Discussion}

We have constructed a nonperturbative framework to analyze one particular
realization of a whole class of nanostructures: a nearly closed system
with ballistic internal dynamics and random losses at the boundary.
Our approach uses a natural regularizer, which enables us to circumvent
the technical difficulties of previous approaches to closed ballistic
systems \cite{Mirlin1,MKhm,AleinerLarkin,EfetovBallistic}. We find
that the resulting theory can be characterized by diffusive modes
confined to the boundary and interacting nonlocally with the interior
as encapsulated in our main result (Eq. (\ref{FreeEnergyfinal})).
These diffusive modes are identified as WGM, which are exponentially
long-lived compared to other trajectories scattering off the boundary
non-tangentially and, therefore are anticipated to dominate the long-time
behavior of response functions. We expect our formalism to be useful
whenever a ballistic nanostructure supports such modes. Note that
no RMT assumptions have been made, and our approach is applicable
in a broad energy range and includes the spatial dependence of physical
quantities.

We derive our results by treating a ballistic billiard as a scattering
system. In particular, we use an effective non-Hermitian Hamiltonian,
which is a common tool in quantum chaotic scattering. However the
resulting supersymmetric NL$\sigma $M is fully intended for the calculations
of {}``internal'' characteristics, such as local density of states
statistic or correlators of eigenfunctions, rather than statistics
of reflection or transmission. For that reason, the escape rate is
chosen to be small and distributed along the boundary. In addition,
we chose the coupling part of the non-Hermitian Hamiltonian to be
fluctuating, thus defining an ensemble of billiards with different
escape rates. These steps allow to study ballistic nanostructures
without the introduction of any other regularizer \cite{WhiteBook,AleinerLarkin}.
The simplest realization of such model is a circular billiard with
constant parameters, for which the surface diffusion NL$\sigma $M
was obtained in \cite{arxivePaper} using the saddle-point solution.
This billiard is also the classical example of a system supporting
WGM.

In this paper we have extended the results of \cite{arxivePaper}
to show that surface diffusion modes are present in generic, chaotic,
smooth-wall billiard as well. To create an extension to other chaotic
optical and electronic systems, for which WGM are of importance \cite{QuantumCorrals,OpticalCorrals,NockelStone,ResonantMagneticVortices},
one needs to adapt our procedure to the details of the coupling to
the environment. However, for most mesoscopic structures the wall
absorption, which is always present in experiment, can be adequately
modeled by uniformly distributed leads. Furthermore, our model allows
us to add finite number of wide leads if the problem under consideration
involves transmission properties. Other  calculations within a similar
framework can be performed for systems which have disorder concentrated
at the boundary, e.g. quantum corral composed of different atoms or
artificial atoms proposed in \cite{ResonantMagneticVortices} with
the random magnitude of the magnetic flux.

As for the handling the calculation of correlators of physical quantities
one needs to introduce a difference between two energies $\omega $
into the derivation. For the final expression for the free energy
(Eq. (\ref{FreeEnergyfinal})) this means a replacement $\epsilon \rightarrow \epsilon +i\omega /2$.
A good starting point can be the application the perturbative technique
described in Ref. \cite{KravtsovMirlin} to the analysis of the $2$-point
function in a circular cavity. Assuming $Q\left(\theta \right)$ ($\theta $
being the polar angle) to fluctuate weakly near the origin $Q_{0}\Lambda $
\cite{arxivePaper}, one can decompose the supermatrix $Q\left(\theta \right)$
as $T^{-1}\widetilde{Q}\left(\theta \right)T$, where $T$ are angle
independent pseudounitary supermatrices, which are, in turn, parametrized
in terms of off-diagonal supermatrices $W$ \cite{Efetovsbook,KravtsovMirlin}.
The perturbation scheme is based on the expansion of $Q\left(\theta \right)$
around the saddle-point solution ($Q_{0}\Lambda $) in terms of $W$
(see the details in \cite{KravtsovMirlin}).

It is also natural to inquire about connection of our model to the
existing ballistic analogues of the diffusive NL$\sigma $Ms, e.g.
the models proposed in \cite{WhiteBook,Mirlin1,MKhm}. Note that the
zero-dimensional versions of these theories produce RMT, which makes
the answer to this question important for studies of truly ergodic
systems. The construction involving diffuse boundary scattering (Ref.
\cite{Mirlin1}) is essentially different from ours, because in Ref.
\cite{Mirlin1} the electron loses memory after a single boundary
collision, while in our model, the WGM trajectories retain phase coherence
until the electron finally leaves the system. Our treatment is complementary
(only in conceptual sense) to that of Refs. \cite{WhiteBook,MKhm},
in which only modes of $Q$ inside the ballistic dot appear, and the
boundary value of $Q$ is nonfluctuating. The zero-dimensional case
$Q\left(\mathbf{r}\right)=const$ inside the dot ($Q\left(\theta \right)=const$
in our case), i.e. the situation in which only lowest mode contributes,
seems to be the only possible connection between these models and
ours. However, another crucial difference between the two formulations
is worth mentioning. The universal parameter $\tau $ (mean free time
between collisions) in the Muzykantskii-Khmelnitskii $\sigma $-model
(\cite{MKhm}) is angular momentum dependent in our case. In order
to address this issue of zero-dimensional correspondence between our
result and these of Refs. \cite{WhiteBook,MKhm} one needs to carry
out the calculations for correlation function of a chaotic system
at hand using the NL$\sigma $M we derived and compare it to the RMT
result in the appropriate limit. This is a subject of ongoing work
to be presented elsewhere.

Finally, we hope to consider applications of our NL$\sigma $M to
the interacting-electron problem in the future. One of the possible
ways to take the interactions into account in diffusive and ballistic
systems with large dimensionless conductance, is to use a {}``Universal
Hamiltonian'' \cite{univ-ham}, which was shown to be the renormalization
group fixed point for weak interactions \cite{longpaper,mm}. We hope
to extend our analysis to the interacting ballistic case by using
the large-$N$ approach of \cite{longpaper}.

We are grateful to the NSF for partial support under DMR-0311761. 

\appendix

\section{Hubbard-Stratonovich decoupling}

In this Appendix we derive the identity which decouples supervector
variables in the action averaged over $\widetilde{\gamma }_{n}$.

For arbitrary function $V\left(u,z\right)I_{4}$, which does not possess
any supersymmetric structure and for two coordinate dependent supermatrices
$Q\left(u,z\right)$ and $A\left(u,z\right)$ which have the same
supersymmetric structure and, therefore commute, the following generalization
of the decoupling rule can be verified:\begin{equation}
\int DQ\textrm{e}^{-\textrm{Str}\int \int \alpha QVQV-i\beta QVAV}=\textrm{e}^{-\frac{\beta }{4\alpha }^{2}\textrm{Str}\int \int AVAV},\label{HSdecoupling}\end{equation}
where we skipped coordinate dependence on $y_{n}$ and $y'_{n}$ and
corresponding differentials. First of all, we make use of the theorem
due to Parisi-Sourlas-Efetov-Wegner (see, for example, \cite{Haakebook}),
which states, that for any {}``invariant superfunction'' the corresponding
superintegral is equal to its value at the origin, which in our case
is unity. By shifting $Q\mapsto Q+i\delta A,$ and setting $2\alpha \delta -\beta =0$,
we get\begin{align}
-\alpha \int _{C_{n}}QV\int _{C_{n}}QV+i\beta \int _{C_{n}}QV\int _{C_{n}}AV\mapsto  & -\alpha \int _{C_{n}}QV\int _{C_{n}}QV-i\left(2\alpha \delta -\beta \right)\nonumber \\
\times \int _{C_{n}}AV\int _{C_{n}}QV=-\alpha \int \int _{C_{n}}Q & VQV-\frac{\beta }{4\alpha }^{2}\int \int _{C_{n}}AVAV.\label{HSdecouple}
\end{align}

Then, we carry out the Hubbard-Stratonovich transformation by applying
Eq. (\ref{HSdecoupling}) from right to left to each of the terms
in Eq. (\ref{AveragedLd}) (in our case $-\beta ^{2}/4\alpha =x_{n}^{2}v_{n}^{2}/8$
and $V\left(y_{n},y'_{n}\right)=\varphi _{n}\left(y_{n}\right)\varphi _{n}^{*}\left(y'_{n}\right)$).
We choose $\alpha =m^{2}/2$, so that $i\beta =mx_{n}v_{n}/2$ and
get Eq. (\ref{Qintroduced}).

\section{Diffusion term}

Approximating the Green functions $\widetilde{\mathcal{G}}\left(r,R\left(s'\right),s,s'\right)$
and $\mathcal{G}\left(r,R\left(s'\right),s,s'\right)$ according to
Eq. (\ref{GreensFunctionIK1}) and $h$ with its boundary value, we
carry out $r$-integration with the help of the identity\[
\int \frac{I_{\alpha }\left(z\right)I_{\beta }\left(z\right)}{z}dz=z\frac{I'_{\alpha }\left(z\right)I_{\beta }\left(z\right)-I_{\alpha }\left(z\right)I'_{\beta }\left(z\right)}{\alpha ^{2}-\beta ^{2}}\]
and get\begin{align*}
F_{t}\left[\delta Q\right] & =i\left(\frac{m^{2}d}{P}\right)^{2}\sum _{l,n,p,q}\int _{0}^{P}ds'\int _{0}^{P}ds\frac{\widetilde{x}\left(s\right)\widetilde{x}\left(s'\right)g\left(s\right)D_{p,q}\left(\tilde{g}\left(s\right)\right)\lambda _{n}}{f\left(s\right)\left(\lambda _{p}+\lambda _{q}\right)}\\
 & \times \textrm{e}^{i\lambda _{l}\int ^{s}\frac{F\left(\overline{s}\right)}{R\left(\overline{s}\right)}d\overline{s}+i\lambda _{n}\int ^{s'}\frac{F\left(\overline{s}\right)}{R\left(\overline{s}\right)}d\overline{s}+i\left(\lambda _{p}+\lambda _{q}\right)\int _{s'}^{s}\frac{F\left(\overline{s}\right)}{R\left(\overline{s}\right)}d\overline{s}}\delta Q_{l}^{\left(t\right)}\delta Q_{n}^{\left(t\right)},
\end{align*}
where $D_{p,q}=1/\left(\mathfrak{j}_{p,q}/\mathfrak{w}_{p,q}+if\left(s\right)\widetilde{g}\left(s\right)\right)$
with\[
\mathfrak{j}_{p,q}=f^{2}I_{\lambda _{p}F}\left(\tilde{g}\right)I_{\lambda _{q}F}\left(\tilde{g}\right)+\widetilde{g}^{2}I'_{\lambda _{p}F}\left(\tilde{g}\right)I'_{\lambda _{q}F}\left(\tilde{g}\right),\]
\begin{eqnarray*}
\mathfrak{w}_{p,q} & = & I_{\lambda _{p}F}\left(\tilde{g}\right)I'_{\lambda _{q}F}\left(\tilde{g}\right)-I'_{\lambda _{p}F}\left(\tilde{g}\right)I_{\lambda _{q}F}\left(\tilde{g}\right).
\end{eqnarray*}
where we omitted $s$-dependence for brevity in the last two lines;
we continue to do that throughout this Appendix. Integration over
$s$ and $s'$ in the above expression for $F_{t}\left[\delta Q\right]$
produces infinitesimally small result unless $n+p+q=0$, $l=-n$.
Anticipating this averages we concentrate on the evaluation of the
sum over $p$ and $q$ restricted by these two conditions. Later on
one can approximate $F_{t}\left[\delta Q\right]$ by replacing the
result of the summation (together with the rest of the integrand)
with its perimeter averaged value. With this in mind we manipulate
$\sum _{l,n,p,q}$ into a double sum \begin{align*}
\sum _{l,n,p,q}D_{pq}\delta Q_{l}^{\left(t\right)}\delta Q_{n}^{\left(t\right)} & \mapsto \sum _{l,n=-\infty }^{\infty }D_{\left|n\right|,\left|l+n\right|}\delta Q_{l}^{\left(t\right)}\delta Q_{-l}^{\left(t\right)}\\
 & \mapsto 2\sum _{\left|l\right|}\delta Q_{\left|l\right|}^{\left(t\right)}\delta Q_{-\left|l\right|}^{\left(t\right)}\sum _{\left|n\right|}\left(D_{\left|n\right|,\left|n\right|+\left|l\right|}+D_{\left|n\right|,\left|n\right|-\left|l\right|}\right).
\end{align*}
 The sum over $\left|n\right|$ can be performed asymptotically in
the limit: $\tilde{g}\left(s\right)\gg 1$, $f\left(s\right)/\tilde{g}\left(s\right)\sim 1$
\cite{Stewartson}. We convert this sum into the integral over the
new variable \[
\mu =\frac{2\pi \left|n\right|F\left(s\right)}{\widetilde{g}\left(s\right)P\overline{\left(F/R\right)}},\]
use uniform expansion for the Bessel function $I_{\lambda _{\left|n\right|}F\left(s\right)}\left(\lambda _{\left|n\right|}F\left(s\right)/\mu \right)$
\cite{Abramowitz} and expand $\mathfrak{j}_{\left|n\right|,\left|n\right|+\left|l\right|}/\mathfrak{w}_{\left|n\right|+\left|l\right|}$
and $\mathfrak{j}_{\left|n\right|,\left|n\right|-\left|l\right|}/\mathfrak{w}_{\left|n\right|-\left|l\right|}$
to the leading order in $1/\widetilde{g}$ . We have\begin{align}
\sum _{\left|n\right|}\left(D_{\left|n\right|,\left|n\right|+\left|l\right|}+D_{\left|n\right|,\left|n\right|-\left|l\right|}\right)= & \frac{PF}{2\pi }\int _{0}^{\infty }d\mu \frac{\left(1-\mu ^{2}-2\mu ^{4}+\frac{f^{2}}{\widetilde{g}^{2}}\right)\lambda _{n}^{2}}{\widetilde{g}\left(1+\mu ^{2}\right)^{3/2}\left(1+\mu ^{2}+\frac{f^{2}}{\widetilde{g}^{2}}\right)}\nonumber \\
 & -i\frac{PF}{2\pi }\int _{0}^{\infty }d\mu \frac{\left(1+f\widetilde{g}\right)\mu ^{2}\lambda _{n}^{2}}{\widetilde{g}\left(1+\mu ^{2}\right)\left(f^{2}+\widetilde{g}^{2}\left(1+\mu ^{2}\right)\right)^{2}},\label{Diffusionconstant3}
\end{align}
 Then, the first integral with respect to $\mu $ in Eq. (\ref{Diffusionconstant3})
vanishes, while the second one yields $\left(\pi /4f^{4}g\right)\left\{ \left(f^{2}+2g^{2}\right)/\sqrt{f^{2}-g^{2}}-2g\right\} $
after we substitute $\widetilde{g}\left(s\right)\rightarrow -ig\left(s\right)$.
To arrive at Eq. (\ref{Diffusivepart}) we drop some of the terms
which are negligible in the above mentioned limit.


\begin{thebibliography}{10}
\bibitem{PhysNano}1992 \emph{Physics of Nanostructures} ed J H Davies and A R Long (Edinburgh:
SUSSP)
\bibitem{Stockmannbook}St\"{o}ckmann H J 1999 \textit{Quantum Chaos: An Introduction} (Cambridge:
Cambridge University Press)
\bibitem{Berry1}Berry M V 1981 \emph{Eur. J. Phys.} \textbf{2} 91
\bibitem{SCA}Jalabert R A, Baranger H U and Stone A D 1990 \emph{Phys. Rev. Lett.}
\textbf{65} 2442; Lin W A and Jensen R V 1996 \emph{Phys. Rev. B}
\textbf{53} 3638
\bibitem{arxivePaper}Rozhkov I and Murthy G, \emph{Preprint: cond-mat}/0504653
\bibitem{WhiteBook}Altland A, Offer C R and Simons B D 1999 \emph{Supersymmetry and Trace
Formulae}, ed Lerner I V \emph{et al} (Dordrecht: Kluwer)
\bibitem{Stone1}Stone A D in Ref. \cite{PhysNano}
\bibitem{Disordred}Lee P A and Ramakrishnan T V 1985 \emph{Rev. Mod. Phys.} \textbf{57}
287; Kramer B and MacKinnon A 1993 \emph{Rep. Prog. Phys.} \textbf{56}
1469
\bibitem{Efetovsbook}Efetov K \textit{\emph{1997}} \textit{Supersymmetry in disorder and
chaos} (Cambridge: Cambridge University Press)
\bibitem{Mirlin1}Blanter Y M, Mirlin A D and Muzikantskii B A 1998 \emph{Phys. Rev.
Lett.} \textbf{80} 4161; Tripathi V and Khmelnitskii D E 1998 \emph{Phys.
Rev.} B \textbf{58} 1122; Blanter Y M, Mirlin A D and Muzikantskii
B A 2001 \emph{Phys. Rev.} B \textbf{63}, 235315
\bibitem{Marcus}Marcus C M \emph{et al} 1992 \emph{Phys. Rev. Lett.} \textbf{69},
506; Kouwenhoven L P \emph{et al.} 1997 in \emph{Proceedings of NATO
ASI on Mesoscopic Electronic Transport} ed by Sohn L L, Sohn G and
Kouwenhoven L P (Dordrecht: Kluwer) 105
\bibitem{QuantumCorrals}Fiete G A and Heller E J 2003 \emph{Rev. Mod. Phys.} \textbf{75} 933
\bibitem{ScweitersAlfordDelos}Schwieters C D, Alford J A and Delos J B 1996 \emph{Phys. Rev. B}
\textbf{54} 10652
\bibitem{Scattering}Fyodorov Y V and Sommers H J 1997 \emph{J. Math. Phys.} \textbf{38}
1918; Dittes F M 2000 \emph{Phys. Rep.} \textbf{339} 215
\bibitem{VWZ}Verbaarschot J M, Weidenmüller H A and Zirnbauer M R 1985 \emph{Phys.
Rep.} \textbf{129}, 367
\bibitem{NonHermitianRM}Fyodorov Y V and Sommers H J 1996 \emph{JETP Lett.} \textbf{63} 970;
Fyodorov Y V, Khoruzhenko B A and Sommers H J 1997 \emph{Phys. Lett.
A} \textbf{226} 46
\bibitem{Lobkis}Lobkis O I, Rozhkov I S and Weaver R L 2003 \emph{Phys. Rev. Lett.}
\textbf{91} 194101
\bibitem{LossySystems}Lewenkopf C H, M\"{u}ller A and Doron D E 1992 \emph{Phys. Rev. A}
\textbf{45} 2635; Pichugin K, Schanz H and Seba P 2001 \emph{Phys.
Rev. E} \textbf{64} 056227; Mendez-Sanchez R A \emph{et al} 2003 \emph{Phys.
Rev. Lett.} \textbf{91} 174102
\bibitem{OpticalCorrals}des Francs G C \emph{et al} 2001 \emph{Phys. Rev. Lett.} \textbf{86}
4950 
\bibitem{NockelStone}N\"{o}ckel J U and Stone A D 1997 \emph{Nature} \textbf{385}, 45;
Turechi H E \emph{et al}. \emph{Preprint: physics}/0308016
\bibitem{ResonantMagneticVortices}Decanini Y and Folacci A 2003 \emph{Phys. Rev. A} \textbf{67} 042704
\bibitem{longpaper}Murthy G \emph{et al} 2004 \emph{Phys. Rev. B} \textbf{69} 075321
\bibitem{MKhm}Muzykantskii B A and Khmelnitskii D E 1995 \emph{Pis'ma Zh. Eksp.
Teor. Fiz.} \textbf{62} 68 {[}\emph{JETP Lett.} 1995 \textbf{62} 76{]};
Andreev A V, Simons B D, Agam O, and Altshuler B L 1996 \emph{Nucl.
Phys. B} \textbf{482} 536
\bibitem{AleinerLarkin}Aleiner I L and Larkin A I 1996 \emph{Phys. Rev. B} \textbf{54} 14423
\bibitem{EfetovBallistic}Efetov K B and Kogan V R 2003 \emph{Phys. Rev. B} \textbf{67} 245312;
Efetov K B, Schwiete G and Takahashi K 2004 \emph{Phys. Rev. Lett.}
\textbf{92} 026807 
\bibitem{Zaitsev1}Zaitsev O, Narevich R and Prange R E 2001 \emph{Foundations of Physics}
\textbf{31} 7 
\bibitem{OpenSystems}Mello P A and Baranger H U 1999 \emph{Waves in Random Media} \textbf{9}
105; Alhassid Y 2000 \emph{Rev. Mod. Phys.} \textbf{72} 895
\bibitem{ThickWire}Zirnbauer M R 1992 \emph{Phys. Rev. Lett.} \textbf{69} 1584
\bibitem{footnote2}Chapter 9 of Ref. \cite{Efetovsbook}
\bibitem{Mine}Rozhkov I, Fyodorov Y V and Weaver R L 2003 \emph{Phys. Rev. E} \textbf{68}
016204; 2004 \emph{ibid}. \textbf{69} 036206
\bibitem{Mirlin}Mirlin A D 2000 \emph{Phys. Rep.} \textbf{326} 259
\bibitem{Stewartson}Stewartson K and Waechter R T 1971 \emph{Proc. Camb. Phil. Soc.} \textbf{69}
353; Sieber M \emph{et al} 1995 \emph{J. Phys. A} \textbf{28} 5041
\bibitem{Abramowitz}Abramowitz M and Stegun I A 1965 \emph{Handbook of mathematical functions}
(New York: Dover)
\bibitem{univ-ham}Brouwer P W, Oreg Y and Halperin B I 1999 \emph{Phys. Rev. B} R13977;
Baranger H U Ullmo D and Glazman L I 2000 \emph{ibid}. \textbf{61}
R2425; Kurland I L, Aleiner I L and Altshuler B.L. 2000 \emph{ibid}.
\textbf{62} 14886
\bibitem{mm}Murthy G and Mathur H 2002 \emph{Phys. Rev. Lett.} \textbf{89} 126804
\bibitem{Haakebook}Haake F 2001 \emph{Quantum Signatures of Chaos} 2nd ed. (Berlin: Springer-Verlag)
\bibitem{KravtsovMirlin}Kravtsov V E and Mirlin A D 1994 \emph{JETP Lett.} \textbf{60} 656\end{thebibliography}
\end{document}